\documentclass[aps, prb, reprint, superscriptaddress, amsmath, amssymb, longbibliography ]{revtex4-2}

\usepackage{graphicx}
\usepackage{bm}
\usepackage{booktabs}
\usepackage{array}
\usepackage{dcolumn}
\usepackage{mathtools}
\usepackage{xcolor}
\definecolor{RED}{rgb}{1,0,0} 
\newif\ifshowrevisions
\showrevisionstrue

\usepackage{hyperref}
\usepackage{microtype}
\usepackage{dsfont}
\usepackage{float}

\hypersetup{
  colorlinks=true,
  linkcolor=blue,
  citecolor=blue,
  urlcolor=blue
}

\newcommand{\kk}{\mathbf{k}}
\newcommand{\qq}{\mathbf{q}}

\newcommand{\sgn}{\mathrm{sgn}}
\newcommand{\FS}{\mathrm{FS}}
\newcommand{\Tc}{T_{\mathrm c}}

\newcommand{\RPA}{\mathrm{RPA}}

\newcommand{\splusminus}{s_{\pm}}

\newcommand{\figplaceholder}[2]{%
\fbox{\begin{minipage}[c][#1][c]{0.96\linewidth}
\centering
\textbf{Figure placeholder}\\[2mm]
#2
\end{minipage}}}

\begin{document}


\title{Pairing-channel reconstruction by Ising spin-orbit coupling in a multiorbital triangular-lattice model}

	\author{Lichuan Wang}
	\affiliation{Anhui Provincial Key Laboratory of Low-Energy Quantum Materials and Devices, High Magnetic Field Laboratory, HFIPS, Chinese Academy of Sciences, Hefei, Anhui 230031, China}
	\affiliation{Science Island Branch of Graduate School, University of Science and Technology of China, Hefei, Anhui 230026, China}
	\author{Yuhang Xiao}
	\affiliation{Anhui Provincial Key Laboratory of Low-Energy Quantum Materials and Devices, High Magnetic Field Laboratory, HFIPS, Chinese Academy of Sciences, Hefei, Anhui 230031, China}
	\affiliation{Science Island Branch of Graduate School, University of Science and Technology of China, Hefei, Anhui 230026, China}
	\author{Ran Wang}
	\affiliation{Anhui Provincial Key Laboratory of Low-Energy Quantum Materials and Devices, High Magnetic Field Laboratory, HFIPS, Chinese Academy of Sciences, Hefei, Anhui 230031, China}
	\affiliation{Science Island Branch of Graduate School, University of Science and Technology of China, Hefei, Anhui 230026, China}
	\author{Xinliang Huang}
	\affiliation{Anhui Provincial Key Laboratory of Low-Energy Quantum Materials and Devices, High Magnetic Field Laboratory, HFIPS, Chinese Academy of Sciences, Hefei, Anhui 230031, China}
	\affiliation{Science Island Branch of Graduate School, University of Science and Technology of China, Hefei, Anhui 230026, China}
	\author{Ning Hao}
	\email{haon@hmfl.ac.cn}
	\affiliation{Anhui Provincial Key Laboratory of Low-Energy Quantum Materials and Devices, High Magnetic Field Laboratory, HFIPS, Chinese Academy of Sciences, Hefei, Anhui 230031, China}
       \affiliation{Collaborative Innovation Center of Advanced Microstructures, Nanjing University, Nanjing 210093, China}

\date{\today}


\begin{abstract}
Ising spin-orbit coupling (SOC) is widely known for locking electronic spins and protecting superconductivity against in-plane paramagnetic pair breaking. Here we show that, in a correlated multiorbital system, Ising SOC can play a qualitatively different role: it reconstructs the hierarchy of spin-fluctuation-mediated pairing channels by reshaping the low-energy Fermi surfaces, orbital-spin coherence factors, and momentum-resolved scattering phase space. We study a noncentrosymmetric two-orbital triangular-lattice model with Rashba SOC, Ising SOC, and onsite SOC. We solve the pairing problem within a multiorbital random-phase-approximation framework. In the single-orbital limit, an $E$-representation state with a dominant $S_z=0$ component emerges only at relatively strong Ising SOC. Orbital degrees of freedom facilitate the stabilization of this pairing state at a smaller nominal Ising SOC strength and generate a broad competition between $E$ and $A_1$ pairing channels. The $E$ state can maintain the same $S_z=0$ triplet character, whereas the $A_1$ state becomes a fully gapped sign-changing $\splusminus$ superconductor. The latter also realizes a topologically nontrivial phase in the time-reversal-invariant weak-pairing regime, with a nonzero $\mathbb{Z}_2$ index, Wilson-loop spectral flow, and boundary modes. Its topology is not carried by a single orbital pairing component, but emerges cooperatively from the dominant intra- and inter-orbital pairing components. These results establish a microscopic route by which Ising SOC and orbital hybridization can reconstruct interaction-driven superconductivity, producing both Ising-triplet and topological singlet-dominated phases in correlated non-centrosymmetric systems.
\end{abstract}

\maketitle


\section{Introduction}
\label{sec:intro}

Spin-orbit coupling (SOC) has become a central ingredient in the modern theory of unconventional superconductivity. In systems without inversion symmetry, anti-symmetric SOC lifts the spin degeneracy of the normal-state bands and entangles spin, momentum, and orbital degrees of freedom~\cite{DAS2021110758,dutta2026coexistencerashbaisingspinsinglet,10.1021/accountsmr.1c00068,PhysRevB.111.014507}. The superconducting order parameter is then generally described by mixed spin-singlet and spin-triplet components, while both the spin response and the pairing interaction can differ qualitatively from those of centrosymmetric superconductors~\cite{PhysRevLett.87.037004,Frigeri_2004,SAMOKHIN20092385,Sato_2017,Sohn2018}. This interplay has been extensively investigated in noncentrosymmetric superconductors and has provided a broad route toward unconventional and topological pairing states~\cite{RevModPhys.82.3045,RevModPhys.83.1057,PhysRevB.93.180501,He2018,PhysRevB.94.094515,PhysRevB.94.060501,PhysRevResearch.2.013026}.

Two-dimensional transition-metal dichalcogenides (TMDs) provide a particularly transparent realization of strong SOC effects. In atomically thin or locally non-centrosymmetric TMD layers, inversion-symmetry breaking permits a predominantly out-of-plane spin-orbit field that changes sign between time-reversal-related valleys\cite{xiao2012coupled}. This so-called Ising SOC produces spin-valley locking and strongly suppresses in-plane paramagnetic pair breaking, leading to upper critical fields well above the conventional Pauli limit~\cite{Xi2016,Saito2016,delaBarrera2018,Zhang_2021,doi:10.1073/pnas.1716781115,Xi2016,doi:10.1126/science.aab2277,mockli2020ising}. Beyond this well-established field-protection effect, Ising SOC can also influence the internal structure of Cooper pairing itself. Previous theoretical studies have shown that SOC, multi-valley Fermi surfaces, electronic correlations, and magnetic fluctuations can generate sizable triplet correlations, unconventional nodal states, and topological superconductivity in TMD-related systems~\cite{PhysRevB.93.180501,He2018,Hsu2017,PhysRevX.10.041003,Divilov_2021,Siegl2025,6rc6-mm3b}.

A less explored question is whether Ising SOC can qualitatively reconstruct the \emph{interaction-driven pairing hierarchy}, rather than merely modify the spin susceptibility of an already established superconducting state. This issue becomes especially important in correlated multi-band or multi-orbital systems. When several orbital components participate near the Fermi level, the pairing kernel depends not only on Fermi-surface geometry but also on momentum-dependent spin-orbital coherence factors\cite{graser2009near}. SOC can then affect superconductivity through several intertwined routes: it changes the low-energy dispersions, redistributes orbital weight, modifies the particle-hole response, and alters the projection of the interaction into distinct Cooper channels. Consequently, varying SOC can reorganize the relative stability of different irreducible representations rather than simply enhance or suppress a fixed pairing state.

This perspective is particularly natural on triangular and hexagonal lattices, where multiple Fermi-surface sheets, low-velocity sectors, van-Hove-proximate structures, and competing finite- and small-momentum scattering processes can occur within a narrow energy window. Weak- and intermediate-coupling studies of TMD-related models have emphasized the importance of electronic correlations, magnetic fluctuations, and Fermi-surface topology in selecting unconventional pairing states~\cite{PhysRevX.10.041003,Divilov_2021,PhysRevB.111.014507,6rc6-mm3b}. Multi-orbital effects add another layer of structure because orbital hybridization and SOC modify the coherence factors entering both the susceptibility and the pairing vertex. More generally, spin-orbital-entangled pairing has been shown to provide a route to unconventional and topological superconductivity in correlated noncentrosymmetric systems~\cite{lpq9-g11z,PhysRevB.110.134517}. Unlike the earlier two-orbital topological-pairing study~\cite{lpq9-g11z}, the present work focuses on the reordering of competing pairing instabilities by Ising SOC.

Motivated by these considerations, we study a minimal two-orbital triangular-lattice model with Rashba SOC, Ising SOC, and an orbital SOC term. The interaction is chosen as a local intra-orbital Hubbard repulsion so that the role of SOC-induced band and coherence-factor reconstruction can be isolated as clearly as possible. We treat the particle-hole fluctuations within the multi-orbital random-phase approximation (RPA) and determine the leading superconducting instabilities from the corresponding linearized Eliashberg equation~\cite{BICKERS1989206,PhysRevB.43.8044,PhysRevB.69.104504}. Our aim is not to construct a quantitative model for a specific material, but to identify generic mechanisms by which Ising SOC and orbital degrees of freedom reorganize the competition among superconducting channels.

We find that the effect of Ising SOC is substantially richer than spin locking alone. In a single-orbital Rashba-Ising model, an $E$-representation state dominated by the $\sigma_z i\sigma_y$ triplet structure can be stabilized, but, within the present benchmark model, only at relatively strong Ising SOC and after the Fermi surface enters a regime of enhanced low-momentum scattering among the small pockets. In the two-orbital model, the inclusion of orbital degrees of freedom enriches the pairing landscape, allowing a triplet-dominated $E$ state to compete with the $A_1$ pairing channel. The $E$ state can become dominated by the $\tau_0\sigma_z i\sigma_y$ component, corresponding to a $d_z$-type, $S_z=0$ triplet configuration. By contrast, the $A_1$ state is predominantly spin singlet and develops a fully gapped sign-changing $\splusminus$ structure.

More remarkably, the $A_1$ state can realize a time-reversal-invariant topological superconducting phase. Its nontrivial topology is not associated with a single pairing component. Instead, it emerges from the cooperation between the dominant $\tau_0\sigma_0 i\sigma_y$ and $\tau_x\sigma_0 i\sigma_y$ components: their superposition removes the nodes of the former and simultaneously reconstructs the projected gap signs on the Fermi surfaces so that the weak-pairing $\mathbb{Z}_2$ criterion becomes nontrivial. The resulting phase is confirmed independently by the Wilson-loop spectral flow and boundary-state calculations.

The coexistence of a triplet-dominated $E$ sector and a topological singlet-dominated $A_1$ sector illustrates a broader principle: Ising SOC can act as a \emph{pairing-channel reconstruction mechanism} in correlated multi-orbital systems. Recent high-field observations in a TMD superlattice have reported a superconducting phase consistent with triplet pairing~\cite{Zhao2026}; this experiment motivates our discussion but does not determine the specific $S_z=0$ order parameter studied here.

The rest of the paper is organized as follows. Section~\ref{sec:model} introduces the model and symmetry structure. Section~\ref{sec:fluct} analyzes the SOC- and orbital-induced reconstruction of magnetic fluctuations and Fermi-surface scattering. Section~\ref{sec:phase} presents the superconducting phase diagram, the evolution of the leading pairing components, and their quasiparticle and spin-response signatures. Details of the multiorbital RPA formalism and pairing-symmetry classification are provided in Secs.~\ref{app:rpa} and \ref{sup:class} of the Supplemental Material, respectively; benchmark calculations are presented in Sec.~\ref{supp:benchmark}.

\section{Model and symmetry}
\label{sec:model}

\subsection{Normal-state Hamiltonian}

We consider a triangular lattice with primitive vectors
\begin{equation}
\mathbf{a}_1=(1,0),\qquad
\mathbf{a}_2=\left(\frac12,\frac{\sqrt3}{2}\right).
\nonumber
\end{equation}
The electronic basis is
\begin{equation}
\Psi_{\kk}
=
\left(
c_{1\kk\uparrow},
c_{1\kk\downarrow},
c_{2\kk\uparrow},
c_{2\kk\downarrow}
\right)^T,
\nonumber
\end{equation}
where $\tau_\mu$ and $\sigma_\mu$ denote Pauli matrices in orbital and spin spaces, respectively. The orbital basis represents effective Wannier states rather than a rotationally invariant atomic multiplet\cite{huang2026spin}; consequently, the conventional relations among the Kanamori parameters are not generally imposed\cite{kanamori1963electron, georges2013strong}. The normal-state Hamiltonian is written as
\begin{equation}
H_{\rm N}
=
\sum_{\kk}
\Psi_{\kk}^{\dagger}
\mathcal{H}_{\rm N}(\kk)
\Psi_{\kk},
\nonumber
\end{equation}
with Rashba and Ising SOC terms of the form\cite{PhysRevLett.95.146802,PhysRevB.110.134517,huang2026spin,PhysRevB.109.195419,lpq9-g11z}
\begin{align}
\mathcal{H}_{\rm N}(\kk)
={}&
\xi_{\kk}\tau_0\sigma_0
-\lambda_R
\left[
g_R^x(\kk)\sigma_x+
g_R^y(\kk)\sigma_y
\right](\tau_0+\tau_x)
\nonumber\\
&-\lambda_I f_I(\kk)\sigma_z(\tau_0+\tau_x)
+\lambda_{\rm so}\tau_y\sigma_z .
\label{eq:Hnormal}
\end{align}
The nearest-neighbor triangular-lattice dispersion is
\begin{equation}
\xi_{\kk}
=
2t
\left(
\cos k_x+
2\cos\frac{k_x}{2}
\cos\frac{\sqrt3 k_y}{2}
\right)-\mu .
\label{eq:disp}
\nonumber
\end{equation}
Throughout the calculations we set $t=-1$ and measure all energy scales in units of $|t|$.
The Rashba form factors are
\begin{align}
g_R^x(\kk)
&=
-\sqrt3
\cos\frac{k_x}{2}
\sin\frac{\sqrt3 k_y}{2},
\nonumber\\
g_R^y(\kk)
&=
\sin k_x+
\sin\frac{k_x}{2}
\cos\frac{\sqrt3 k_y}{2},
\nonumber
\end{align}
The Ising SOC is described by a triangular-lattice form factor \cite{huang2026spin}
\begin{equation}
f_I(\kk)
=
\sin k_x
-
2\sin\frac{k_x}{2}
\cos\frac{\sqrt3 k_y}{2}.
\label{eq:IsingFF}
\nonumber
\end{equation}
The onsite SOC term,
\begin{equation}
H_{\rm orb-SOC}
=
\lambda_{\rm so}\tau_y\sigma_z
\label{eq:orbSOC}
\end{equation}
is treated as an orbital SOC within the two-orbital subspace. If the two basis orbitals are regarded as an effective doublet carrying a nonzero projected $L_z$, then $\tau_y$ represents the anti-symmetric orbital generator and Eq.~\eqref{eq:orbSOC} is the projected form of an atomic-like $L_zS_z$ coupling. 
A representative set of parameters used in the main text is
$\{\lambda_R,\lambda_I,\lambda_{\rm so},\mu,n\}$
=$\{0.2,0.4,0.4,0.9472, 2.34\}$,
unless otherwise stated.

\begin{figure}[t]
	\centering
	\IfFileExists{fig1.png}{\includegraphics[width=1.0\columnwidth]{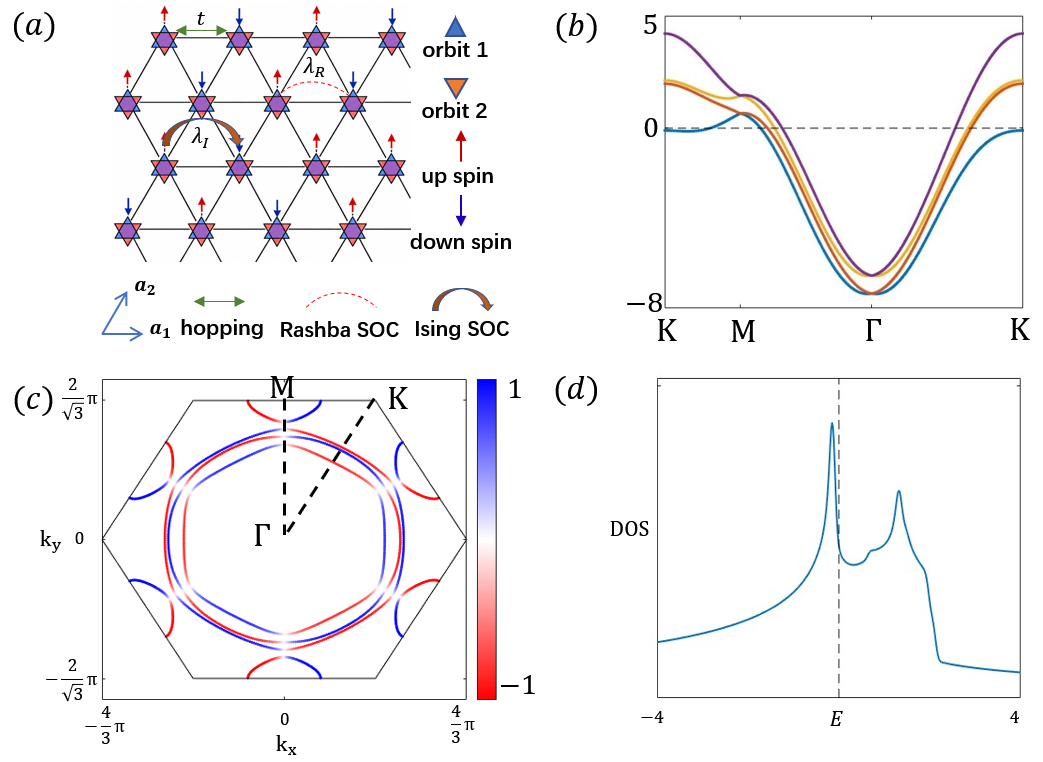}}{\figplaceholder{36mm}{fig1.png (external image not supplied)}}
	\caption{
		Normal-state electronic structure of the effective triangular-lattice model with spin-orbit coupling.
		(a) Real-space lattice setup.
		(b) Band structure along the high-symmetry path \(K-M-\Gamma-K\). The dashed horizontal line denotes the Fermi level.
		(c) Fermi surface in the hexagonal Brillouin zone at the same filling. The dashed lines indicate the high-symmetry path used in (b). The color on the Fermi surface represents the out-of-plane spin polarization \(\langle\sigma_z\rangle\), with blue and red denoting positive and negative values, respectively.
		(d) Density of states as a function of energy. The dashed vertical line marks the Fermi level. In all the plottings, the parameters $\{\lambda_R,\lambda_I,\lambda_{\rm so}\}$
=$\{0.2,0.4,0.4\}$
	}
	\label{fig:normal}
\end{figure}

Figure~\ref{fig:normal} summarizes the normal-state electronic structure. Rashba SOC produces momentum-dependent in-plane spin canting, while the Ising term generates a sizable out-of-plane spin polarization. The onsite orbital SOC further splits and hybridizes the low-energy bands. In the parameter regime relevant to the pairing reconstruction, the Fermi level lies close to a low-velocity regime without being tuned exactly to a singular van Hove point. Consequently, the superconducting instability is controlled by a combination of Fermi-surface geometry, $1/v_F$ weighting, spin texture, and orbital coherence rather than by the density of states alone.

\subsection{Interaction}

The general multi-orbital Hubbard interaction can be expressed as following,
\begin{equation}
	\begin{aligned}
	H_{\rm int}
	&=U\sum_{i \alpha}\hat{n}_{i \alpha\uparrow}\hat{n}_{i \alpha\downarrow}
	+V\sum_{i \alpha<\beta}\sum_{\sigma\sigma'}\hat{n}_{i \alpha \sigma}\hat{n}_{i \beta \sigma'}\\
	&-J\sum_{i, \alpha<\beta}\bm{S}_{i\alpha}\cdot\bm{S}_{i\beta}\\
	&+J'\sum_{i, \alpha<\beta}\sum_{\sigma}c^{\dagger}_{i\alpha\sigma}c^{\dagger}_{i\alpha\bar{\sigma}}
	c_{i\beta\bar{\sigma}}c_{i\beta\sigma}.
\end{aligned}
\label{eq:Hint}
\end{equation}
The interaction is formulated in an effective Wannier-orbital basis rather than in a set of atomic orbitals with well-defined angular momenta. Consequently, the effective intra-orbital Hubbard interaction may incorporate contributions from both intra- and inter-orbital Coulomb interactions in the original atomic basis. We retain only this effective intra-orbital repulsion as a minimal interaction model, neglecting the remaining inter-orbital and exchange terms to isolate the role of spin-orbital reconstruction in the pairing instability, i.e., $V,J,J'\rightarrow 0$.

The superconducting gap is expanded as
\begin{equation}
\Delta(\kk)
=
\sum_{\alpha,\beta}
\Delta_{\alpha\beta}(\kk)
\tau_\alpha\sigma_\beta i\sigma_y .
\nonumber
\label{eq:gapexpand}
\end{equation}
Fermionic anti-symmetry requires
\begin{equation}
\Delta(\kk)=-\Delta^T(-\kk).
\label{eq:fermiasym}
\nonumber
\end{equation}
Because $\tau_{0,x,z}^T=+\tau_{0,x,z}$ whereas $\tau_y^T=-\tau_y$, momentum parity, spin structure, and orbital exchange parity must be treated together. The complete $C_{3v}$ classification used in the numerical analysis is given in Sec.~\ref{sup:class} of the Supplemental Material.


\section{SOC- and orbital-induced reconstruction of magnetic fluctuations}
\label{sec:fluct}

Within the RPA, the effective pairing interaction is inherited from the momentum and matrix structure of the particle-hole susceptibility. A compact summary of the formalism is given in Sec.~\ref{app:rpa} of the Supplemental Material. Here we focus on the physical origin of the pairing reconstruction.
\begin{figure}[t]
	\centering
	\IfFileExists{fig2.png}{\includegraphics[width=1.0\columnwidth]{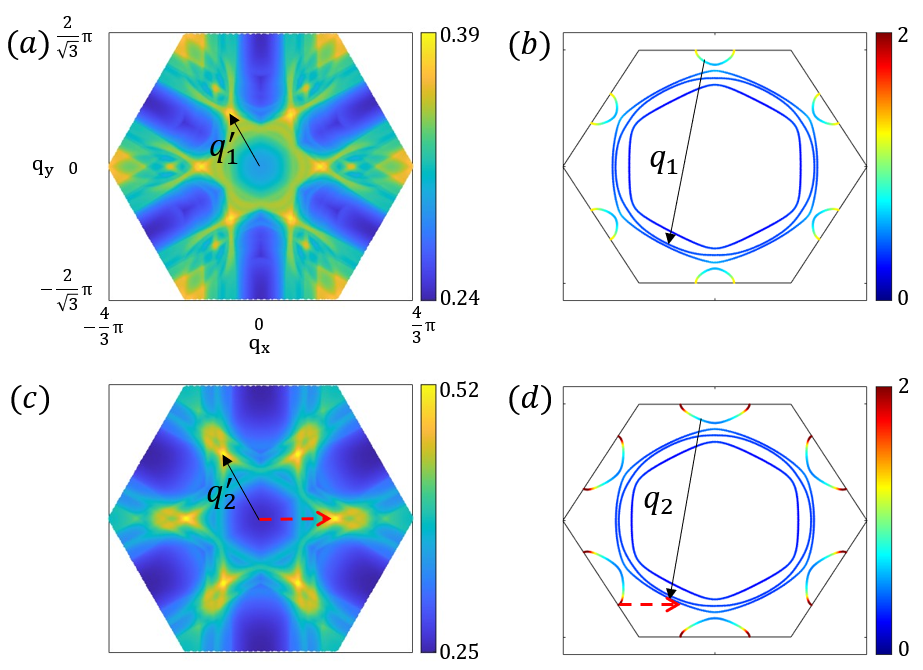}}{\figplaceholder{36mm}{fig2.png (external image not supplied)}}
	\caption{
		Relation between the leading spin-susceptibility response and the Fermi-surface scattering geometry.
		(a,c) Largest eigenvalue of the spin susceptibility matrix in momentum space for
		\(\mu=1.0972\) and \(\mu=0.8972\), respectively.
		(b,d) Corresponding Fermi surfaces colored by the inverse Fermi velocity \(1/v_F\).
		The black arrows indicate susceptibility-peak scattering vectors, whereas the red dashed arrows identify enhanced weighted scattering channels. 
		Momenta differing by a reciprocal lattice vector are equivalent, so
		\(\mathbf q_1\equiv \mathbf q_1'+\mathbf G\) and
		\(\mathbf q_2\equiv \mathbf q_2'+\mathbf G\).
		For \(\mu=1.0972\), the dominant response is associated with nesting between the \(\Gamma\)-centered Fermi surface and the small pockets around the \(M\) points.
		For \(\mu=0.8972\), besides the analogous \(\mathbf q_2\) process, an additional scattering channel appears near the crossing region of the small Fermi pockets, as indicated by the red dashed arrows in panels (c) and (d).
	}
	\label{fig:scattering}
\end{figure}
The change of the leading pairing channel can be understood from the evolution of the dominant response wave vectors, as shown in Fig.~\ref{fig:scattering}. 

To characterize the dominant magnetic fluctuations, we examine the largest eigenvalue of the static spin-susceptibility matrix, following a diagnostic commonly employed in multiorbital spin-fluctuation studies~\cite{PhysRevB.69.104504,PhysRevB.81.054502,PhysRevB.105.134509,PhysRevB.92.104511}. For \(\mu=1.0972\), the strongest response occurs near $\mathbf q_1$, which connects nearly parallel segments of the $\Gamma$-centered Fermi surface and the small pockets around the $M$ points.
This nesting-enhanced scattering generates a strong repulsive interaction between the connected Fermi-surface regions and favors a sign change between them. 
Since the low-energy states in this parameter regime are mainly governed by Ising SOC, with \(\lambda_I>\lambda_R\), this sign-changing structure naturally supports an odd-in-momentum, \(d_z\)-dominated triplet component.
For \(\mu=0.8972\), the response still contains the analogous scattering vector \(\mathbf q_2\), which connects the \(\Gamma\)-centered Fermi surface with the \(M\)-centered pockets. 
However, an additional scattering channel appears near the crossing region of the small \(M\)-centered pockets, as marked by the red dashed arrows in Figs.~\ref{fig:scattering}(c) and \ref{fig:scattering}(d). This channel does not coincide with the dominant susceptibility peaks.
Although this channel is not associated with extended Fermi-surface nesting, its contribution can be enhanced by the large phase-space weight of low-velocity Fermi-surface segments.
\begin{equation}
\mathcal{N}_{\mathrm{FS}}
=
\oint_{\mathrm{FS}}
\frac{d\ell}{|v_F(\mathbf{k})|}.
\label{eq:fs_weight}
\end{equation}
Here, $d\ell$ is a line element along the two-dimensional Fermi surface. Therefore, the regions with small Fermi velocity can dominate the projected kernel even without extended geometric nesting.
Together with the \(\mathbf q_2\) process, it introduces repulsive scattering involving the same \(M\)-pocket region and couples small-pocket segments separated by approximately \(120^\circ\) in the Brillouin zone. 
This frustration favors an effective sign arrangement among the \(M\)-centered pockets, driving the system from the \(\sigma_z\)-dominated odd-in-momentum triplet state toward a singlet-dominated \(s_{\pm}\) state. The red scattering vector in Fig.~\ref{fig:scattering}(d) is identified by maximizing the weighted susceptibility over selected pairs of Fermi-surface patches:
\begin{equation}
\begin{aligned}
(i_*,j_*) &\in
\underset{(i,j)\in\mathcal{C}}
{\operatorname{arg\,max}}
\left[
w_i w_j
\chi_{\max}(\mathbf{k}_i-\mathbf{k}_j)
\right],\\
\mathbf{q}_{\mathrm{red}}
&=
\mathbf{k}_{i_*}-\mathbf{k}_{j_*}.
\end{aligned}
\label{eq:weighted_scattering}
\end{equation}
Here, $\mathcal C$ denotes the selected pairs of patches on the relevant Fermi pockets, and $\chi_{\max}(\mathbf q)$ is the largest eigenvalue of the static spin-susceptibility matrix. The discrete Fermi-surface weight is

\begin{equation}
w_i =
\frac{\Delta\ell_i}{|v_F(\mathbf{k}_i)|}.
\label{eq:discrete_weight}
\end{equation}

\section{Competing superconducting pairing channels}
\label{sec:phase}

\begin{figure}[t]
\centering
	\IfFileExists{fig3.png}{\includegraphics[width=1.0\columnwidth]{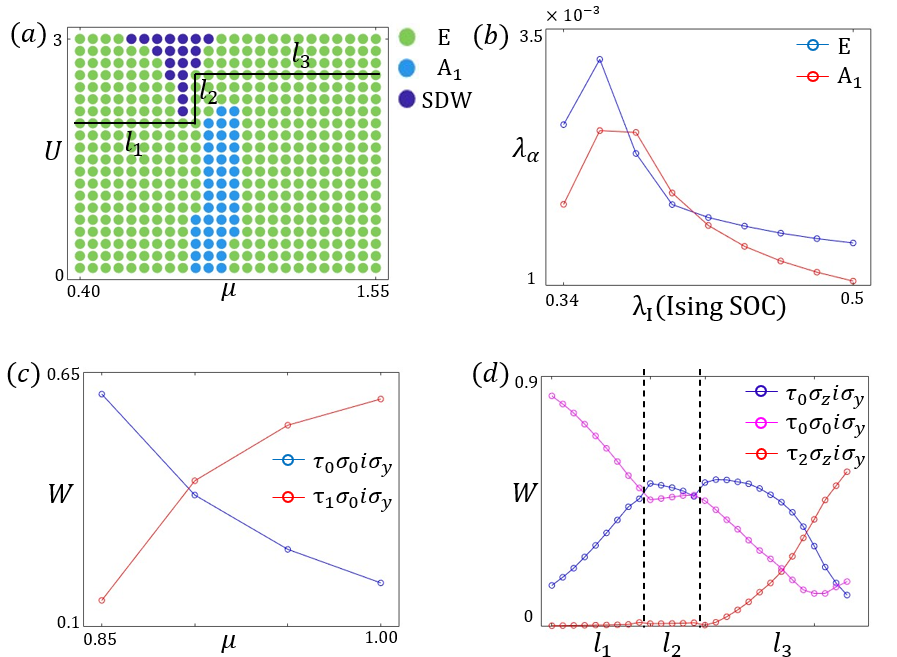}}{\figplaceholder{36mm}{fig3.png (external image not supplied)}}
	\caption{
		Pairing phase diagram and internal channel evolution obtained from the linearized Eliashberg equation.
		(a) \(U-\mu\) phase diagram of the leading order. The purple, blue, and green regions denote the SDW phase, the \(A_1\)-IR superconducting phase, and the \(E\)-IR superconducting phase, respectively. The three paths \(l_1\), \(l_2\), and \(l_3\) indicate the representative cuts used in (d).
		(b) Leading Eliashberg eigenvalues in the \(A_1\)-IR and \(E\)-IR channels as functions of the Ising SOC strength \(\lambda_I\).
		(c) Component weights of the \(\tau_0\sigma_0\) and \(\tau_x\sigma_0\) channels in the \(A_1\)-IR solution with \(U=0.6\).
		(d) Evolution of the dominant components in the \(E\)-IR solution along the combined path \(l_1\)-\(l_2\)-\(l_3\). The dashed vertical lines separate the three path segments.
	}	
	\label{fig:phase}
\end{figure}
\begin{figure*}[t]
	\centering
	\IfFileExists{fig4.png}{\includegraphics[width=2.0\columnwidth]{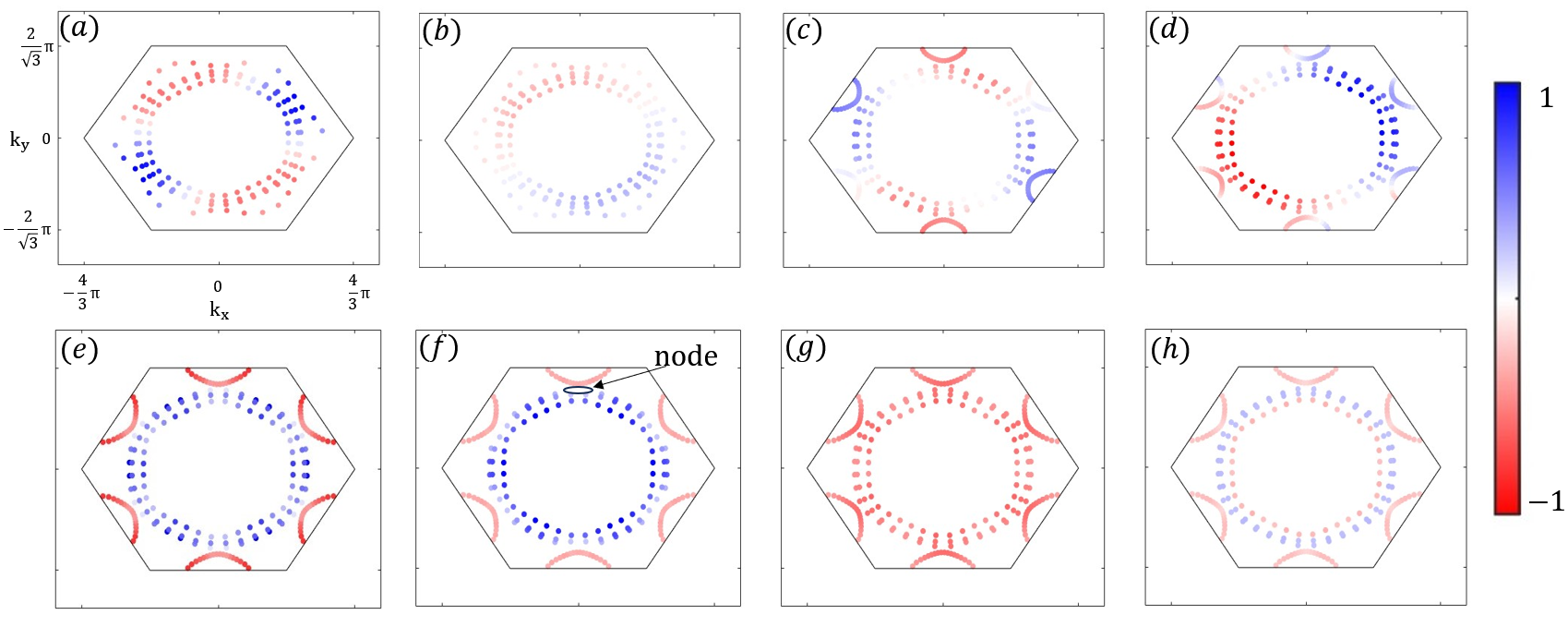}}{\figplaceholder{36mm}{fig4.png (external image not supplied)}}
	\caption{
		Momentum-space pairing structures for representative superconducting instabilities. Panels (a)--(d) display the $\tau_0\sigma_0 i\sigma_y$ and $\tau_0\sigma_z i\sigma_y$ components of two $E$-representation states: (a,b) a singlet-dominated state at $\{\mu,U\}=\{0.4,0.6\}$ and (c,d) a $d_z$-triplet-dominated state at $\{\mu,U\}=\{1.05,0.6\}$. Panels (e)--(h) characterize the $A_1$ $s_{\pm}$ state at $\{\mu,U\}=\{0.9,0.6\}$. (e) Projected pairing-gap signs for the complete $A_1$ gap. (f) Projected gap structure of the $\tau_0\sigma_0 i\sigma_y$ component considered separately; its nodes preclude assigning a fully gapped two-dimensional $\mathbb Z_2$ invariant to this component alone. (g,h) Gap structure and projected pairing-gap signs obtained by retaining only the $\tau_x\sigma_0 i\sigma_y$ component. Red and blue indicate opposite projected-gap signs.
	}
	\label{fig:states}
\end{figure*}
The superconducting instability is obtained from the linearized Eliashberg equation using the full orbital-spin pairing vertex. Figure~\ref{fig:phase}(a) summarizes the leading pairing channels in the $(U,\mu)$ plane.
The $A_1$ region occupies a finite parameter window near a low-$v_F$ regime, while the two-dimensional $E$ representation occupies a broader parameter region. Figure~\ref{fig:phase}(b) shows that the leading $A_1$ eigenvalue is not a monotonic function of $\lambda_I$; instead, it is enhanced only over an intermediate SOC range. This behavior indicates that the $A_1$ instability is not controlled by spin locking alone. It results from a combined reconstruction of Fermi-surface geometry, orbital content, and projected scattering strength.

The dominant $A_1$ components are
$\tau_0\sigma_0 i\sigma_y$
and
$\tau_x\sigma_0 i\sigma_y$.
Their relative weights are shown in Fig.~\ref{fig:phase}(c).
Together, these two components account for approximately $70\%$ of the total pairing weight in the representative topological regime. We define the component weight as
\[
W(\tau_{i}\sigma_{j}i\sigma_{y})=\frac{\sum_{\kk}|\Delta(\tau_{i}\sigma_{j}i\sigma_{y},\kk)|^2}{\sum_{i,j}\sum_{\kk}|\Delta(\tau_{i}\sigma_{j}i\sigma_{y},\kk)|^2}
\] 
The state is therefore spin-singlet dominated, but its orbital structure is intrinsically multi-component.
The $E$ sector exhibits a qualitatively different evolution, as shown in Fig.~\ref{fig:phase}(d). In the phase diagram, along \(l_1\),  the leading solution has a strong $\tau_0\sigma_0 i\sigma_y$ component, whereas along \(l_2\) and part of \(l_3\), the $\tau_0\sigma_z i\sigma_y$ component becomes dominant. The crossover is driven by the changing Fermi-surface scattering phase space and by the increasing importance of the Ising-polarized low-energy states. Table~\ref{class} lists selected pairing basis functions for $C_{3v}$. Other channels make negligible contributions in our numerical solutions. The complete classification is given in Sec.~\ref{sup:class} of the Supplemental Material.

\subsection{$E$ representation: Ising-triplet sector}
\begin{table}
	\caption{Selected pairing basis functions for $C_{3v}$. The function $g^{(n)}(\kk)$ denotes an $A_1$-symmetric form factor for the $n$th-neighbor shell; for example, $g^{(0)}(\kk)=1$ and $g^{(1)}(\kk)=\cos k_x+2\cos\frac{1}{2}k_x\cos \frac{\sqrt{3}}{2}k_y$. $d_1(\kk)=\sqrt{3}\sin \frac{1}{2}k_x\sin \frac{\sqrt{3}}{2}k_y$, $d_2(\kk)=\cos k_x -\cos \frac{1}{2}k_x\cos \frac{\sqrt{3}}{2}k_y$, $\eta_x(\kk)=\sin k_x+\sin \frac{1}{2}k_x\cos \frac{\sqrt{3}}{2}k_y$ and $\eta_y(\kk)=\sqrt{3}\cos \frac{1}{2}k_x\sin \frac{\sqrt{3}}{2}k_y$.}
	\begin{ruledtabular}
		\begin{tabular}{ccc}
			Pairing form & $\phi$ / $\boldsymbol{d}$ &IRs for $C_{3v}$ \\ \hline 
			$i\phi\sigma^2\tau^{0/1(3)}$ & $g^{(n)}(\kk)$ & $A_1(A_2)$ \\
			& $\begin{cases}
				-d_1(\kk) \\
				d_2(\kk)
			\end{cases}$ & $E$ \\
			$i(\bm{d}\cdot\bm{\sigma})\sigma^2\tau^{0/1(3)}$
			& $\begin{cases}
				(0,0,-\eta_y(\kk))\\
				(0,0,\eta_x(\kk))
			\end{cases}$ &\\
		\end{tabular}
	\end{ruledtabular}
	\label{class}
\end{table}
The representative triplet-dominated $E$ solution has the structure
\begin{equation}
\Delta_E(\kk)
\simeq
d_z(\kk)\,
\tau_0\sigma_z i\sigma_y
+
\Delta_E^{(s)}(\kk)\,
\tau_0\sigma_0 i\sigma_y
+\cdots .
\label{eq:Egap}
\nonumber
\end{equation}
The leading triplet component is odd in momentum for the orbital-symmetric $\tau_0$ sector and has $\mathbf{d}\parallel \mathbf{z}$. This is the same spin configuration that is naturally robust against an in-plane Zeeman field in an Ising environment.
Because $E$ is a two-dimensional irreducible representation, with $\Delta(\kk)=\eta_1\Delta_1(\kk)+\eta_2\Delta_2(\kk)$, the linearized Eliashberg equation determines the degenerate basis at $\Tc$ but does not by itself determine the low-temperature combination of the two components. A real combination can produce a nematic nodal state, whereas a complex combination may gap part or all of the nodal structure. Determining the actual state below $\Tc$ requires the quartic free energy or a fully nonlinear gap calculation. The Ginzburg--Landau free energy up to quartic order\cite{sigrist1991phenomenological} is
\begin{equation}
	\begin{aligned}
		F&=\alpha(|\eta_1|^2+|\eta_2|^2)+\beta_1(|\eta_1|^2+|\eta_2|^2)^2\\
		&+\beta_2|\eta_1\eta_2^*-\eta_2\eta_1^*|^2+O(\eta^6)
	\end{aligned}
	\nonumber
\end{equation}
Writing $(\eta_1,\eta_2)=\Delta_0(\cos\theta,\sin\theta\,e^{i\phi})$, we obtain
\[
F=\alpha|\Delta_0|^2+\beta_{eff}(\theta, \phi)|\Delta_0|^4 +O(\eta^6)
\]
where \(\beta_{eff}(\theta, \phi)=\beta_1+\beta_2\sin^2 2\theta \sin^2 \phi\). The minimum free energy below the transition is
\[
F_{min}=-\frac{\alpha^2}{4\beta_{eff}}
\]
For $\beta_2>0$, the free energy favors a real order-parameter configuration, corresponding to a nematic superconducting state. For $\beta_2<0$, a chiral combination of the two components is favored, spontaneously breaking time-reversal symmetry, $\Delta(\kk)\propto\Delta_1(\kk)+i\Delta_2(\kk)$.
We therefore restrict the present analysis to the pairing instability and its dominant spin-orbital structure and do not assign a bulk topological invariant to the $E$ state. Representative numerical gap harmonics for the $E$ representation are
\begin{align}
	\Delta_{E,s}^{(1)}(\kk)
	\simeq{}&
	0.07d_1(\kk)-0.04
	d_2(\kk),
	\nonumber
	\\
	\Delta_{E,t}^{(1)}(\kk)
	\simeq{}&
	0.014\eta_x(\kk)
	-0.025\eta_y(\kk),
	\nonumber
	\\
	\Delta_{E,s}^{(2)}(\kk)
	\simeq{}&
	-0.016d_1(\kk)-0.027d_2(\kk),
	\nonumber
	\\
	\Delta_{E,t}^{(2)}(\kk)
	\simeq{}&
	0.062\eta_x(\kk)
	+0.036\eta_y(\kk).
	\nonumber
\end{align}
\subsection{$A_1$ representation: orbital-cooperative topological \texorpdfstring{$s_{\pm}$}{s+-} superconductivity}

\begin{figure}[t]
	\centering
	\IfFileExists{fig5.png}{\includegraphics[width=1.0\columnwidth]{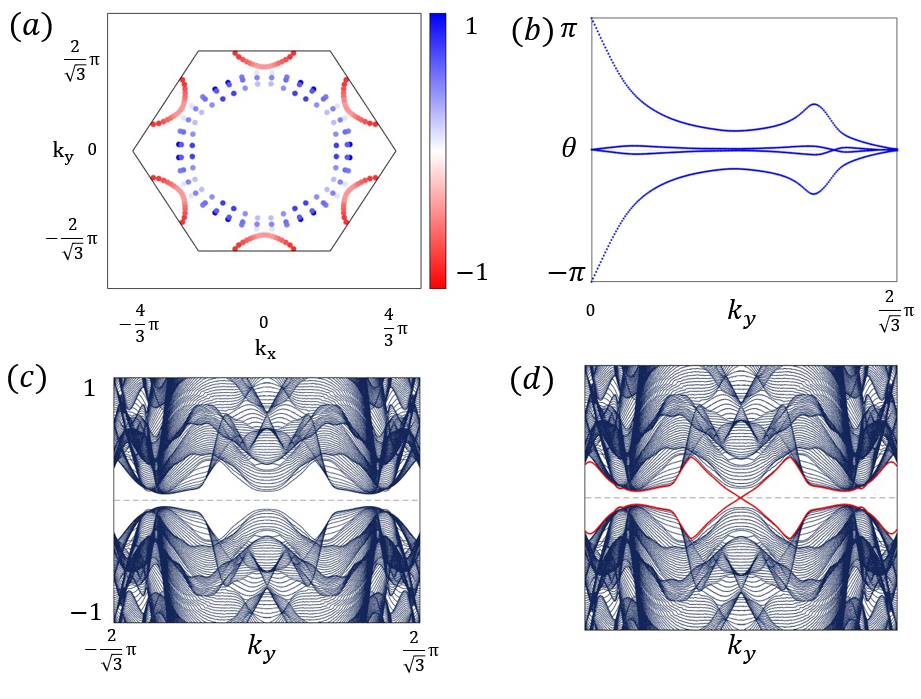}}{\figplaceholder{36mm}{fig5.png (external image not supplied)}}
	\caption{
		Topological characterization of the \(A_1\)-IR \(s_\pm\) superconducting state.
		(a) Topological criterion on the Fermi surface for a representative \(A_1\)-IR solution. The color indicates the sign and magnitude of the projected pairing amplitude, revealing an $s_\pm$-type sign reversal between different Fermi-surface sectors.
		(b) Wilson-loop spectrum~\cite{PhysRevB.84.075119} of the occupied BdG bands. Its nontrivial spectral flow is consistent with the $\mathbb{Z}_2=1$ invariant.
		(c) Quasiparticle spectrum in a reference geometry, showing the bulk superconducting gap of the representative \(A_1\)-IR state.
		(d) Boundary spectrum in a slab geometry. Gapless boundary modes appear inside the bulk superconducting gap, consistent with the nontrivial bulk topology.
	}
	\label{fig:topology}
\end{figure}
The $A_1$ state is fully gapped in the representative topological regime and is dominated by two spin-singlet orbital components:
\begin{equation}
\Delta_{A_1}(\kk)
\simeq
\Delta_0(\kk)\,
\tau_0\sigma_0 i\sigma_y
+
\Delta_x(\kk)\,
\tau_x\sigma_0 i\sigma_y.
\label{eq:A1gap}
\nonumber
\end{equation}
Comparing the dominant orbital components reveals their cooperative role in the topological state. The $\tau_0\sigma_0 i\sigma_y$ component by itself is nodal. The $\tau_x\sigma_0 i\sigma_y$ component by itself is topologically trivial. Their superposition removes the nodes and simultaneously reconstructs the projected pairing sign on the non-degenerate Fermi surfaces.

The representative harmonic decompositions extracted from the numerical solution are
\begin{align}
\Delta_0(\kk)
\simeq{}&
0.04+0.03g^{(1)}(\kk)-0.026g^{(3)}(\kk),
\nonumber\\
\Delta_x(\kk)
\simeq{}&
-0.035
+0.0092
g^{(2)}(\kk).
\nonumber
\end{align}
These expressions are used only to illustrate the momentum structure of the dominant numerical eigenfunction; the topology is evaluated from the complete BdG Hamiltonian. The topological assignment is corroborated by Wilson-loop spectral flow and gapless boundary modes in a slab geometry, as shown in Fig. \ref{fig:topology}.

For a fully gapped time-reversal-invariant weak-pairing superconductor, the Fermi-surface criterion~\cite{PhysRevB.81.134508} can be expressed as
\begin{equation}
(-1)^\nu
=
\prod_j
\left[
\sgn(\delta_j)
\right]^{m_j},
\label{eq:Z2FS}
\end{equation}
where
\begin{equation}
\delta_j(\kk)
=
\langle u_{j\kk}|
\mathcal{T}\Delta^\dagger(\kk)
|u_{j\kk}\rangle
\nonumber
\end{equation}
is the projected pairing amplitude on Fermi surface $j$, and $m_j$ is the parity of the number of time-reversal-invariant momenta enclosed by that Fermi surface~\cite{PhysRevB.81.134508}. We note that \(\langle u_{j\kk}|\mathcal{T}f(\kk)\tau_0\sigma_0 i\sigma_y|u_{j\kk}\rangle=f(\kk)\). For the representative $A_1$ state, Eq.~\eqref{eq:Z2FS} gives $\nu=1$.

\subsection{Quasi-particle and spin-response signatures}
\label{sec:response}
\begin{figure}[t]
	\centering
	\IfFileExists{fig6.png}{\includegraphics[width=1.0\columnwidth]{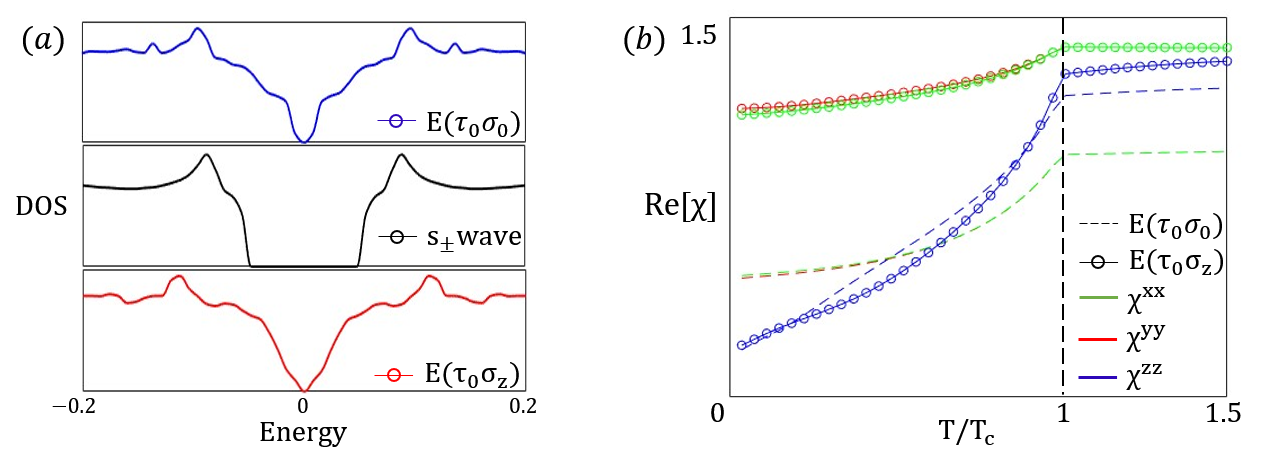}}{\figplaceholder{36mm}{fig6.png (external image not supplied)}}
	\caption{
		BdG quasi-particle and spin-response properties of representative superconducting states.
		(a) Superconducting quasiparticle density of states (DOS) for the singlet-dominated E state (top), the topological $A_1$ $s_\pm$ state (middle), and the $d_z$-triplet-dominated E state (bottom). The curves are vertically offset for clarity.
		(b) Normalized Knight shift for different magnetic-field orientations.
	}
	\label{fig:response}
\end{figure}
The representative $A_1$ and $E$ states have distinct low-energy and spin-response fingerprints. The fully gapped $\splusminus$ state exhibits a U-shaped quasi-particle density of states, whereas the representative real $E$ basis states have stronger low-energy spectral weight associated with nodes or deep anisotropy.

The Knight shift provides a particularly direct probe of the internal spin structure. The \(\tau_0\sigma_0i\sigma_{y}\)-dominated $E$ state shows a strong suppression of the superconducting spin susceptibility in all directions, modulo SOC-induced anisotropy. In contrast, the $d_z$-dominated $E$ state retains a much larger transverse response. This behavior follows from the predominantly $S_z=0$ triplet structure and provides a possible way to distinguish the two pairing sectors.


\section{Discussion and conclusions}
\label{sec:discussion}

The present results reveal a hierarchy of mechanisms. Ising SOC first produces an out-of-plane spin texture, but the superconducting consequences are not determined by spin locking alone. SOC also changes the band energies, Fermi velocities, and orbital-spin wave functions. These changes reconstruct the particle-hole susceptibility and the matrix elements entering the RPA Cooper vertex:
\begin{equation}
\lambda_I,\lambda_{\rm so}
\Rightarrow
\{E_n(\kk),|u_{n\kk}\rangle\}
\Rightarrow
\chi(\qq)
\Rightarrow
V_{\rm pair}(\kk,\kk')
\Rightarrow
\Delta(\kk).
\label{eq:mechanismchain}
\nonumber
\end{equation}
This is the sense in which Ising SOC reconstructs rather than merely protects superconductivity.

The topological $A_1$ phase illustrates a second consequence of orbital reconstruction. The nontrivial topology is not associated with a pure odd-parity state, nor with a single orbital component. Instead, the full projected gap changes sign among time-reversal-invariant Fermi surfaces in a way that satisfies the $\mathbb{Z}_2$ weak-pairing criterion. The cooperation of the two leading spin-singlet orbital components is essential both to remove nodes and to generate the nontrivial projected sign structure.

Our model is intended as a generic microscopic framework rather than a material-specific description. The recent observation of a high-field superconducting phase in a BaTa$_2$S$_5$ superlattice~\cite{Zhao2026} provides motivation for triplet superconductivity in this setting, but a quantitative treatment of that material would require the actual superlattice folding, realistic multiorbital band structure, and magnetic-field-dependent pairing kernel. The present result instead establishes that correlated multiorbital Ising systems provide a natural interaction-based route to both $d_z$-triplet and topological $\splusminus$ superconductivity.

A remaining issue concerns the internal order-parameter selection within the two-dimensional $E$ representation. The linearized theory determines the instability at $\Tc$ but not whether the low-temperature state is nematic or chiral. This requires a quartic Ginzburg-Landau analysis or a self-consistent nonlinear gap calculation and is left for future work.

We have investigated spin-fluctuation-mediated superconductivity in a noncentrosymmetric two-orbital triangular-lattice model incorporating Rashba, Ising, and orbital SOC. We demonstrate that Ising SOC can actively reconstruct the pairing hierarchy, rather than merely protect superconductivity against paramagnetic pair breaking, through its interplay with orbital degrees of freedom and Fermi-surface scattering. This reconstruction gives rise to competing $E$ and $A_1$ pairing instabilities. The $E$ state can be dominated by the $S_z=0$ triplet component $\tau_0\sigma_z i\sigma_y$, whereas the $A_1$ state realizes a fully gapped topological $s_\pm$ superconductor with $\mathbb{Z}_2=1$. The nontrivial topology of the latter emerges from the cooperation between the leading intra- and inter-orbital singlet pairing components, $\tau_0\sigma_0 i\sigma_y$ and $\tau_x\sigma_0 i\sigma_y$. Our results establish a microscopic mechanism by which Ising SOC and orbital degrees of freedom jointly reshape superconducting pairing symmetry and topology in correlated noncentrosymmetric systems.

\begin{acknowledgments}
This work was supported by the National Key R\&D Program of China (Grants No. 2022YFA1403200 and No.2024YFA1613200), National Natural Science Foundation of
China (Grants No. 92265104, No. 12022413), the Basic Research Program of the Chinese
Academy of Sciences Based on Major Scientific Infrastructures (Grant No. JZHKYPT-2021-08), the CASHIPS Directors Fund (Grant No. BJPY2023A09), Anhui Provincial Major S\&T
Project(s202305a12020005), and the High Magnetic Field Laboratory of Anhui Province under
Contract No. AHHM-FX-2020-02
\end{acknowledgments}

\bibliographystyle{apsrev4-2}
\bibliography{ref}


\clearpage
\onecolumngrid

\setcounter{section}{0}
\renewcommand{\thesection}{S\arabic{section}}
\renewcommand{\thesubsection}{S\arabic{section}.\arabic{subsection}}
\renewcommand{\theequation}{S\arabic{equation}}
\renewcommand{\thefigure}{S\arabic{figure}}
\renewcommand{\thetable}{S\arabic{table}}
\setcounter{equation}{0}
\setcounter{figure}{0}
\setcounter{table}{0}

\begin{center}
{\large\bf Supplemental Material for}\\[1mm]
{\large\bf ``Pairing-channel reconstruction by Ising spin-orbit coupling in a multiorbital triangular-lattice model''}
\end{center}

%
\section{Multiorbital RPA formalism}
\label{app:rpa}

We define the generalized bare susceptibility in orbital-spin space as
\begin{align}
	\chi^{l_1\sigma_1,l_2\sigma_2}_{l_3\sigma_3,l_4\sigma_4}
	(\qq,i\nu_n)
	={}&
	-\frac{T}{N}
	\sum_{\kk,i\omega_n}
	G_{l_3\sigma_3,l_1\sigma_1}(\kk,i\omega_n)
	\nonumber
	\times
	G_{l_2\sigma_2,l_4\sigma_4}
	(\kk+\qq,i\omega_n+i\nu_n),
	\label{eq:chi0Matsu}
\end{align}
with
\begin{equation}
	\hat G(\kk,i\omega_n)
	=
	\left[
	i\omega_n-\mathcal{H}_{\rm N}(\kk)
	\right]^{-1}.
	\nonumber
\end{equation}
In the orbit-spin representation, the static susceptibility is
\begin{align}
	\chi^{l_1\sigma_1,l_2\sigma_2}_{l_3\sigma_3,l_4\sigma_4}
	(\qq)
	={}&
	-\frac{1}{N}
	\sum_{\kk,\mu\nu}
	a_{\mu}^{l_3\sigma_3}(\kk)
	a_{\mu}^{l_1\sigma_1*}(\kk)
	\times
	a_{\nu}^{l_2\sigma_2}(\kk+\qq)
	a_{\nu}^{l_4\sigma_4*}(\kk+\qq)
	\nonumber
	\times
	\frac{
		f[E_\nu(\kk+\qq)]-f[E_\mu(\kk)]
	}{
		E_\nu(\kk+\qq)-E_\mu(\kk)+i0^+
	}.
	\label{eq:chiBand}
\end{align}
\begin{equation}
	\begin{aligned}
		&W^{l_1\sigma l_1\bar{\sigma}}_{l_1\sigma l_1\bar{\sigma}}=U && W^{l_1\sigma l_1\sigma}_{l_1\bar{\sigma} l_1\bar{\sigma}}=-U\\
		&W^{l_2\sigma l_1\bar{\sigma}}_{l_2\sigma l_1\bar{\sigma}}=V && W^{l_1\sigma l_1\sigma}_{l_2\bar{\sigma} l_2\bar{\sigma}}=-V\\
		&W^{l_1\sigma l_1\bar{\sigma}}_{l_2\sigma l_2\bar{\sigma}}=J && W^{l_1\sigma l_2\sigma}_{l_1\bar{\sigma} l_2\bar{\sigma}}=-J\\
		&W^{l_1\sigma l_2\bar{\sigma}}_{l_1\sigma l_2\bar{\sigma}}=J' && W^{l_1\sigma l_2\sigma}_{l_2\bar{\sigma} l_1\bar{\sigma}}=-J'\\
	\end{aligned}
	\nonumber
\end{equation}
For the intraorbital Hubbard interaction, the nonzero spin-orbital interaction matrix elements are chosen consistently with Eq.~\eqref{eq:Hint}. The RPA susceptibility is
\begin{equation}
	\hat\chi^{\RPA}(\qq)
	=
	\left[
	\hat 1-\hat\chi^0(\qq)\hat W
	\right]^{-1}
	\hat\chi^0(\qq).
	\label{eq:chiRPA}
	\nonumber
\end{equation}

The effective Cooper interaction contains the bare interaction together with ladder and bubble fluctuation corrections:
\begin{equation}
	\hat V(\kk,\kk')
	=
	\hat W
	+
	\hat V_l(\kk,\kk')
	-
	\hat V_b(\kk,\kk'),
	\label{eq:Vfull}
	\nonumber
\end{equation}
where
\begin{align}
	\left[V_b\right]^{\,l_1\sigma_1,l_2\sigma_2}_{l_3\sigma_3,l_4\sigma_4}
	(\kk,\kk')
	&=
	\left[
	\hat W\hat\chi^{\RPA}(\kk-\kk')\hat W
	\right]^{
		l_1\sigma_1,l_4\sigma_4
	}_{l_3\sigma_3,l_2\sigma_2},
	\nonumber\\
	\left[V_l\right]^{\,l_1\sigma_1,l_2\sigma_2}_{l_3\sigma_3,l_4\sigma_4}
	(\kk,\kk')
	&=
	\left[
	\hat W\hat\chi^{\RPA}(\kk+\kk')\hat W
	\right]^{
		l_1\sigma_1,l_2\sigma_2
	}_{l_3\sigma_3,l_4\sigma_4}.
\nonumber
\end{align}

The linearized Eliashberg equation is
\begin{align}
	\lambda\,
	\Delta_{l_1\sigma_1,l_4\sigma_4}(\kk)
	={}&
	-T
	\sum_{\kk',i\omega_n}
	\sum_{l_2l_3\sigma_2\sigma_3}
	V^{\,l_1\sigma_1,l_2\sigma_2}_{l_3\sigma_3,l_4\sigma_4}
	(\kk,\kk')
	\nonumber
	\times
	F_{l_3\sigma_3,l_2\sigma_2}
	(\kk',i\omega_n),
	\label{eq:Eliashberg}
	\nonumber
\end{align}
with
\begin{align}
	F(\kk,i\omega_n)
	=
	G(\kk,i\omega_n)
	\Delta(\kk)
	G^T(-\kk,-i\omega_n).
	\nonumber
\end{align}
The leading eigenvalue $\lambda$ determines the dominant superconducting instability.
\section{\texorpdfstring{Pairing-symmetry classification for $C_{3v}$}{Pairing-symmetry classification for C3v}}
\label{sup:class}
\begin{table}[H]
	\caption{Pairing basis functions classified by irreducible representations of $C_{3v}$.}
	\begin{ruledtabular}
		\begin{tabular}{ccc}
			Pairing form & $\phi$ / $\boldsymbol{d}$ &IRs for $C_{3v}$ \\ \hline 
			$i\phi\sigma^2\tau^{0/1(3)}$ & $\mathds{1}$ & $A_1(A_2)$ \\
			& $\begin{cases}
				-d_1 \\
				d_2
			\end{cases}$ & $E$ \\
			$i(\bm{d}\cdot\bm{\sigma})\sigma^2\tau^{0/1(3)}$ & $(-\eta_y,\eta_x,0)$ & $A_1(A_2)$ \\
			& $\begin{cases}
				(k^3,0,0) \\
				(0,k^3,0)
			\end{cases}$ & $E$ \\
			& $\begin{cases}
				(0,0,-\eta_y)\\
				(0,0,\eta_x)
			\end{cases}$ & \\
			& $\begin{cases}
				(2\eta_x-k^3,2\eta_y,0)\\
				(2\eta_y,3\sin k_x,0)
			\end{cases}$ & \\
			$i\phi\sigma^2\tau^2 $ & $k^3$ & $A_1$ \\
			& $\begin{cases}
				-\eta_y \\
				\eta_x
			\end{cases}$ & $E$ \\
			$i(\bm{d}\cdot\bm{\sigma})\sigma^2\tau^2$ & $(0,0,\mathds{1})$&$A_1$ \\
			& $\begin{cases}
				(\mathds{1},0,0)\\
				(0,\mathds{1},0)
			\end{cases}$ & $E$ \\
			& $\begin{cases}
				(0,0,-d_1) \\
				(0,0,d_2)
			\end{cases}$ &\\
			& $\begin{cases}
				(-3\cos k_x,2d_1,0) \\
				(2d_1,g(k),0)
			\end{cases}$ &\\
			$i\phi\sigma^2\tau^{3}$
			& $\begin{cases}
				d_2\\
				d_1 
			\end{cases}$ & $E$ \\
			$i(\bm{d}\cdot\bm{\sigma})\sigma^2\tau^{3}$ & $(\eta_x,\eta_y,0)$ & $A_1$ \\
			& $(0,0,k^3)$ &\\
			& $\begin{cases}
				(0,k^3,0)\\
				(-k^3,0,0)
			\end{cases}$ & $E$ \\
			& $\begin{cases}	
				(0,0,\eta_x)\\
				(0,0,\eta_y)
			\end{cases}$ &\\
			& $\begin{cases}	
				(2\eta_y,k^3-2\eta_x,0)\\
				(3\sin k_x, -2\eta_y)
			\end{cases}$ &\\
		\end{tabular}
	\end{ruledtabular}
\end{table}
Here $\mathds{1}$ denotes an $A_1$-symmetric form factor, for example $1$ or $\cos k_x+2\cos\frac{k_x}{2}\cos\frac{\sqrt{3}k_y}{2}$. $d_1=\sqrt{3}\sin \frac{1}{2}k_x\sin \frac{\sqrt{3}}{2}k_y$, $d_2=\cos k_x -\cos \frac{1}{2}k_x\cos \frac{\sqrt{3}}{2}k_y$, $k^3=\sin k_x-2\sin \frac{1}{2}k_x\cos \frac{\sqrt{3}}{2}k_y$, $\eta_x=\sin k_x+\sin \frac{1}{2}k_x\cos \frac{\sqrt{3}}{2}k_y$, $\eta_y=\sqrt{3}\cos \frac{1}{2}k_x\sin \frac{\sqrt{3}}{2}k_y$ and $g(k)=\cos k_x-4\cos\frac{1}{2}k_x\cos \frac{\sqrt{3}}{2}k_y$.
\section{\texorpdfstring{Benchmark calculations for the single- and two-orbital models}{Benchmark calculations for the single- and two-orbital models}}
\label{supp:benchmark}
\subsection{Single-orbital benchmark}

We first consider the single-orbital Rashba-Ising model with representative parameters
\begin{equation}
\{\lambda_{\rm so},\lambda_R,\lambda_I\}
=
\{0,0.2,0.8\}.
\nonumber
\end{equation}
As the chemical potential crosses the Fermi-surface topology transition, the system evolves from two $\Gamma$-centered sheets to one $\Gamma$-centered sheet plus three boundary-centered pockets. The superconducting region is predominantly in the two-dimensional $E$ representation, while sufficiently strong $U$ drives an SDW instability.
\begin{figure}[h]
\centering
	\IfFileExists{figA1.png}{\includegraphics[width=0.8\columnwidth]{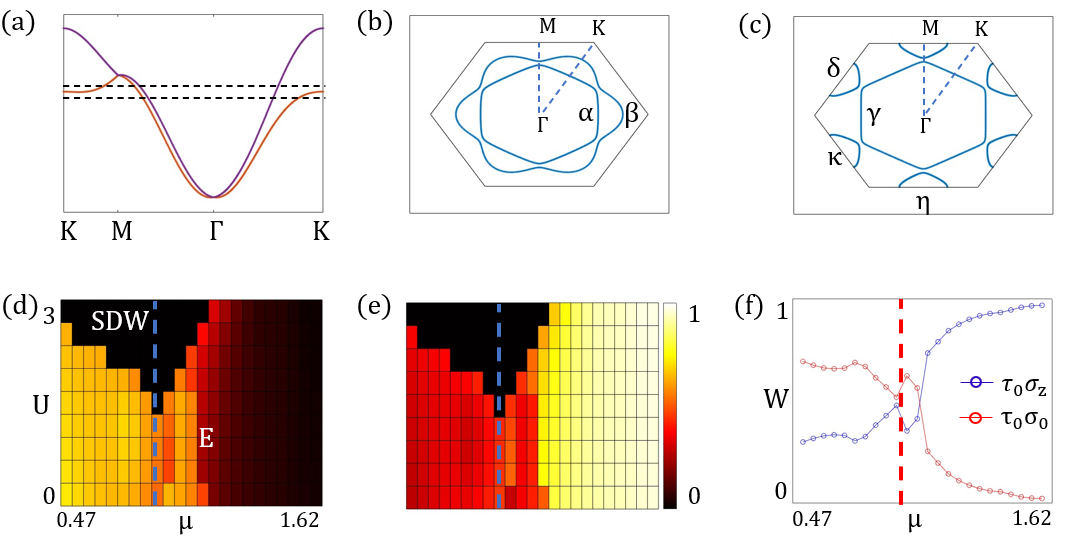}}{\figplaceholder{36mm}{figA1.png (external image not supplied)}}
	\caption{
		Single-orbital Rashba-Ising model and the evolution of its leading \(E\)-representation pairing state.
		(a) Band structure along the high-symmetry path \(K-M-\Gamma-K\). The two dashed horizontal lines denote the chemical potentials corresponding to the Fermi surfaces in (b) and (c).
		(b,c) Fermi surfaces inside the first Brillouin zone. The solid black hexagon marks the Brillouin-zone boundary, the dashed lines indicate the high-symmetry path, and the blue curves denote the Fermi surfaces. Panel (b) contains two \(\Gamma\)-centered sheets (Fermi surface \(\alpha\), \(\beta\)), whereas panel (c) contains one \(\Gamma\)-centered sheet (Fermi surface \(\gamma\)) and three \(M\)-centered pockets (Fermi surface \(\delta, \kappa, \eta\)).
		(d,e) Relative weights of the
		\(\sigma^0 i\sigma_y\) and
		\(\sigma^z i\sigma_y\) components in the leading \(E\)-representation gap.
		Brighter colors indicate larger weights.
		The blue dashed line marks the Fermi-surface topology transition between the structures shown in (b) and (c). The black regions in the upper left of panels (d) and (e) represent SDW order, while the remaining parameter space is occupied by the $E$-representation superconducting instability.
		(f) Weight of \(\tau_0\sigma_0\) and \(\tau_0\sigma_z\) components with the \(\tau_0\sigma_0\) and \(\tau_0\sigma_z\) marked in red and blue, respectively.}
	\label{fig:Ssingle}
\end{figure}
Panels (d)--(f) of Fig.~\ref{fig:Ssingle} show that the Fermi-surface transition does not immediately generate a dominant $\sigma_z i\sigma_y$ component. Immediately beyond the transition, the leading $E$ state still contains a sizable $\sigma_0 i\sigma_y$ contribution. The triplet-dominated regime appears only after a further evolution of the Fermi-surface scattering phase space.

The scattering measure used in the original benchmark calculation is
\begin{equation}
S_{ab}
=
\frac{
\sum_{i\in\FS_a}
\sum_{j\in\FS_b}
w_iw_j
\Theta[\chi_{\kk_i-\kk_j}-\chi_p]
}{
p
\sum_{i\in\FS_a}
\sum_{j\in\FS_b}
w_iw_j
},
\nonumber
\end{equation}
The triplet-dominated regime is characterized by an enhancement of intrapocket and small-$q$ scattering on the three small Fermi surfaces.
\begin{table}[H]
	\caption{Relative scattering strength between different Fermi-surface sheets for representative chemical potentials. \(\chi_{p=0.05}=(1-p) \times max\{eig[\chi(q)]\}\) which selects susceptibility values within $5\%$ of the maximum. Zero entries indicate the absence of scattering between the specified Fermi-surface sheets with susceptibility exceeding $\chi_p$.}
	\begin{ruledtabular}
		\begin{tabular}{ccccc}
			Chemical potential& $\gamma-\gamma$ &$\gamma-\delta$ & $\delta-\delta$ & $\delta-\eta$\\ \hline 
			\addlinespace
			0.9213 & 1.74 & 2.63 & 0 & 2.12\\
			\addlinespace
			1.0713 & 1.61 & 4.13 & 0.46 & 0\\
			\addlinespace
			1.1213 & 1.40 & 3.67 & 0.90 & 0\\
			\addlinespace
			1.1713 & 1.07 & 2.75 & 2.00 & 0
		\end{tabular}
	\end{ruledtabular}
\end{table}

\subsection{Two-orbital model in the degenerate limit}
We first set $\lambda_{\rm so}=0$ while retaining the remaining SOC terms. In this limit, the $\tau_x=-1$ orbital combination is unaffected by the Rashba and Ising terms and forms a spin-degenerate low-energy band. The pairing phase diagram is then qualitatively different from that of the non-degenerate model: the $A_1$ region expands, while the $E$ and SDW sectors are strongly reduced.
\begin{figure}[h]
\centering
\IfFileExists{figA2.png}{\includegraphics[width=1\columnwidth]{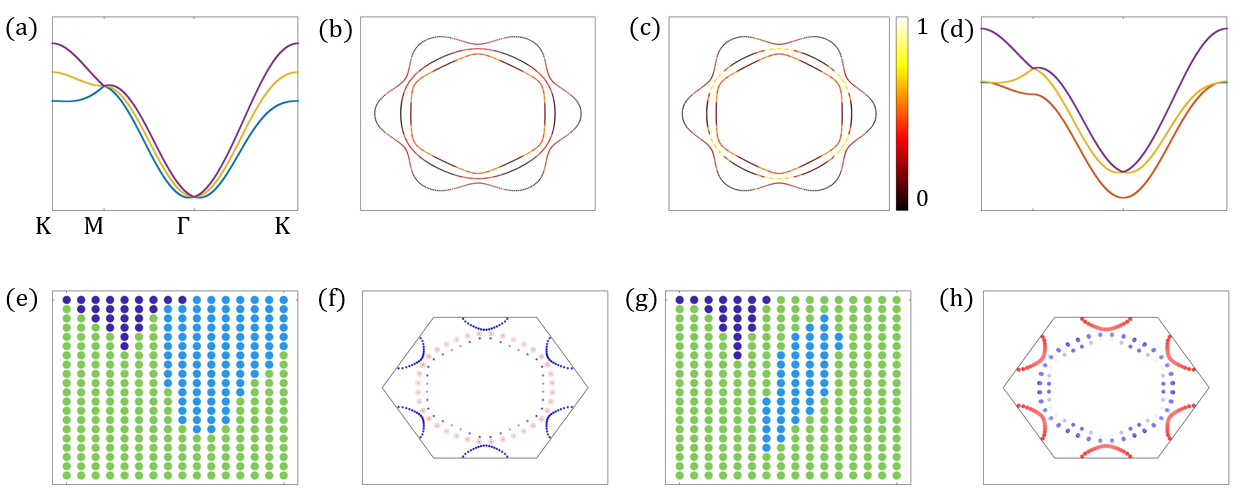}}{\figplaceholder{36mm}{figA2.png (external image not supplied)}}
\caption{
	Electronic structure, pairing phase diagrams, and scattering properties of the two-orbital model as the onsite orbital SOC varies.
	(a) Band structure along the high-symmetry path \(K-M-\Gamma-K\). The yellow band remains doubly degenerate.
	(b) Scattering accumulation on the individual degenerate Fermi surfaces.
	(c) Combined scattering accumulation after summing over the two degenerate Fermi surfaces.
	(d) Band structure after adding an allowed inter-orbital spin-independent hopping term proportional to \(\tau^x\sigma^0\). The band degeneracy is not lifted by this term.
	(e) Superconducting phase diagram in the \((\mu,U)\) plane with \(\lambda_{\rm so}=0.04\). The \(A_1\) representation occupies a large portion of the phase space, while the \(E\) and SDW regions are reduced compared with the single-orbital case.
	(f) Topological criterion of the leading \(A_1\) superconducting state. Red and blue denote opposite gap signs, while filled and open symbols distinguish the two nearly degenerate Fermi-surface sheets at finite $\lambda_{\rm so}$. The resulting sign structure is topologically trivial.
	(g) Pairing phase diagrams for \(\lambda_{\mathrm{so}}=0.24\).
	A fully gapped topological $s_{\pm}$ state emerges for these parameters; its projected pairing-gap signs at $\mu=0.97$ are shown in (h). The projected gap signs satisfy the nontrivial weak-pairing Fermi-surface criterion.
}
\label{fig:Sdeg}
\end{figure}
The two degenerate bands do not merely double the susceptibility magnitude. Their superposition changes where the largest scattering weight occurs. A spin-independent inter-orbital term proportional to $\tau_x\sigma_0$ does not remove the relevant degeneracy, showing that generic hybridization and orbital SOC play distinct roles.
Increasing the onsite orbital SOC suppresses the degeneracy-driven enhancement of the $A_1$ channel and restores the $E$ representation over a broad parameter range. Upon further increasing $\lambda_{\rm so}$, the low-energy band splitting and orbital coherence factors are reconstructed sufficiently to generate the fully gapped topological $A_1$ $\splusminus$ state.

\section{Fermi-surface-resolved scattering across chemical potential}

The Fermi-surface-resolved scattering matrices are evaluated for chemical potentials $0.8472\leq\mu\leq1.0972$.
In the multiorbital analysis, the Fermi-surface patch weight $w_i$ is defined by Eq.~\eqref{eq:discrete_weight}.
Strong finite-momentum scattering between the three $\Gamma$-centered sheets and the three small boundary pockets is present over a broad parameter window. At lower chemical potential, the low-$v_F$ sectors contribute an additional enhancement.
\begin{table}[H]
	\caption{Relative scattering strength between different Fermi-surface sheets for representative chemical potentials. \(\chi_p=0.95 \times max\{eig[\chi(q)]\}\). Zero entries indicate the absence of inter-sheet scattering with susceptibility exceeding $\chi_p$.}
	\begin{ruledtabular}
		\begin{tabular}{cccccccccccc}
			Chemical potential& $1-1$ &$2-2$ & $3-3$ & $1-2$ & $1-3$ & $2-3$ & $4-4$ & $1-4$ & $2-4$ & $3-4$ & $4-5$\\ \hline 
			\addlinespace
			0.8472 & 0.14 & 0.15 & 0.05 & 0.30 & 0.34 & 0.27 & 0.1 & 0.86 & 0.8 & 0.48 & 0.14\\
			\addlinespace
			0.8972 & 0.19 & 0.13 & 0.03 & 0.34 & 0.22 & 0.15 & 0.32 & 0.73 & 0.69 & 0.44 & 0\\
			\addlinespace
			0.9472 & 0.17 & 0.10 & 0.03 & 0.24 & 0.18 & 0.12 & 0.33 & 0.63 & 0.58 & 0.42 & 0\\
			\addlinespace
			0.9972 & 0.15 & 0.09 & 0.05 & 0.19 & 0.18 & 0.14 & 0.23 & 0.63 & 0.60 & 0.36 & 0\\
			\addlinespace
			1.0472 & 0.15 & 0.10 & 0.11 & 0.22 & 0.20 & 0.17 & 0 & 0.52 & 0.58 & 0.38 & 0\\
			\addlinespace
			1.0972 & 0.16 & 0.11 & 0.12 & 0.23 & 0.22 & 0.18 & 0 & 0.50 & 0.58 & 0.36 & 0\\
		\end{tabular}
	\end{ruledtabular}
\end{table}

\clearpage
\end{document}